\documentclass[aps,pra,10pt,twocolumn,groupedaddress,notitlepage,showpacs,floatfix]{revtex4-2}
\pdfoutput=1

\usepackage{graphicx,graphics,epsfig,subfigure,times,bm,bbm,amssymb,amsmath,amsfonts,amsthm,mathrsfs,MnSymbol}
\usepackage{gensymb}
\usepackage[matrix,frame,arrow]{xypic}
\usepackage[pdfstartview=FitH]{hyperref}
\usepackage{subfigure}
\hypersetup{
    colorlinks=true,       
    linkcolor=red,          
   citecolor=magenta,        
    filecolor=magenta,      
    urlcolor=cyan,           
    runcolor=cyan
}
\usepackage[pdftex]{color}
\usepackage{braket}
\usepackage{enumerate}
\usepackage[normalem]{ulem}
\usepackage[usenames,dvipsnames]{xcolor}
\usepackage{multirow}
\usepackage{mathtools}

\definecolor{orange}{rgb}{1,0.5,0}

\usepackage{tikz}

\newcommand{\bes} {\begin{subequations}}
\newcommand{\ees} {\end{subequations}}
\newcommand{\bea} {\begin{eqnarray}}
\newcommand{\eea} {\end{eqnarray}}

\definecolor{gold}{rgb}{0.85,.66,0}

\newcommand{\beq}{\begin{equation}}

\newcommand{\eeq}{\end{equation}}

\newcommand{\ignore}[1]{}

\def\s{\sigma}

\def\>{\rangle}
\def\<{\langle}

\def\s0{I}

\newcommand{\ig}[1]{}

\begin{document}

\bibliographystyle{unsrt}

\date{\today}

\title{Transforming photon statistics through zero-photon subtraction}
\author{C. M. Nunn}
\author{S. U. Shringarpure}
\author{T. B. Pittman}
\affiliation{Physics Department, University of Maryland Baltimore
County, Baltimore, MD 21250}

\begin{abstract}

Zero-photon subtraction (ZPS) is a conditional measurement process that can reduce the mean photon number of quantum optical states without physically removing any photons.  Here we show that ZPS can also be used to transform certain super-Poissonian states into sub-Poissonian states, and vice versa.  Combined with a well-known ``no-go'' theorem on conditional measurements, this effect leads to a new set of non-classicality criteria that can be experimentally tested through ZPS measurements. 
\end{abstract}


\maketitle

\section{Introduction}\label{sec:intro}

The investigation of super-Poissonian and sub-Poissonian light sources plays a significant role in the history of quantum optics~\cite{hbt1956interferometer,short1983observation}. The photon-number distributions of these sources display variances that are, respectively, wider or narrower than the benchmark Poissonian statistics of a coherent state with the same average photon number~\cite{agarwal2012quantumoptics}.  Experimentally, these properties can be conveniently characterized by Mandel's $Q$ parameter, with $Q > 0$ for super-Poissonian sources and ${-1 \leq Q < 0}$ for sub-Poissonian sources~\cite{mandel1979subpoissonian}.

Given either type of source, it is also interesting to consider the physical processes that can be used to actively transform the emitted super-Poissonian light into sub-Poissonian light, and vice versa. Quintessential examples include the use of optical nonlinearities such as two-photon absorption~\cite{gilles1993tpa} or photon blockades~\cite{birnbaum2005photonblockade} to transform super- into sub-Poissonian light, and amplification~\cite{hong1985conditions,boyd2008propagation} or phase randomization \cite{li2020groundglass} to transform sub- into super-Possonian light. More recently, it has been shown that conditional measurement processes such as photon addition and photon subtraction can also be used to implement these statistical transformations~\cite{agarwal1991nonclassical,ourjoumtsev2006kittens,zavatta2008subtracting,barnett2018statistics}. Here we show that the relatively new conditional measurement process of zero-photon subtraction (ZPS) can also transform certain sub-Poissonian states into super-Poissonian states (and vice versa), despite no photons being added to or subtracted from the system. Other related work includes performing statistical transformations by manipulating the initial parameters in atomic fluorescence~\cite{arnoldus1983conditions}, sideband squeezing~\cite{grosse2007antibunching}, photon catalysis~\cite{bartley2012multiphoton}, and displacement operation~\cite{deoliveira1990properties} experiments.

\begin{figure}[t]
\includegraphics[scale=0.45,trim={270 180 270 155},clip]{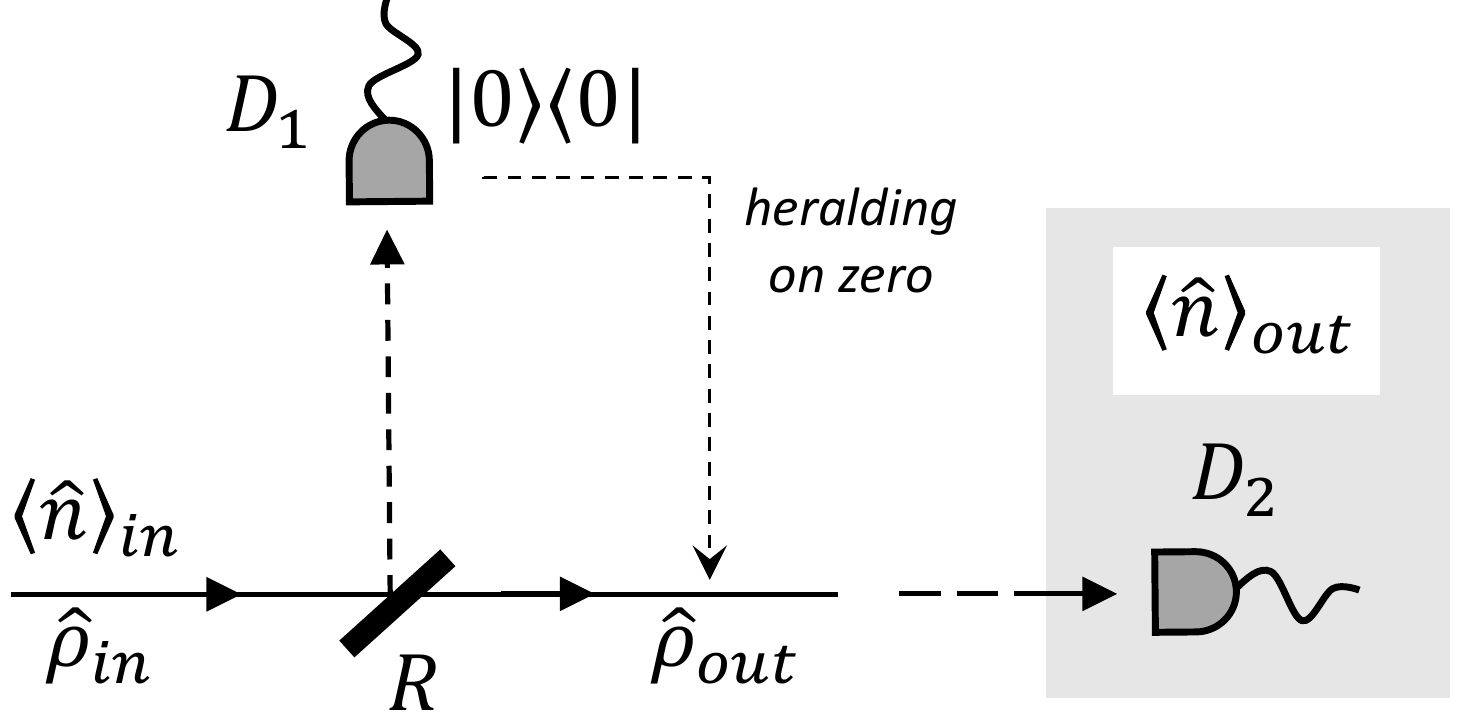}
\caption{Overview of a typical zero-photon subtraction (ZPS) setup.  An input state $\hat{\rho}_{in}$ enters one input port of a beamsplitter with variable reflectivity $R$, and the detection of zero reflected photons at $D_{1}$ heralds a noiselessly attenuated output state $\hat{\rho}_{out}$. Despite no photons being removed from the system, ZPS can dramatically transform the photon statistics of certain input states. These effects can be observed and quantified by measuring trends in a relative attenuation parameter $K(R)$ using an auxiliary detector $D_{2}$.}
\label{fig:overview}
\end{figure}

Figure~\ref{fig:overview} shows an overview of a typical ZPS setup. ZPS was originally proposed in the context of noiseless attenuation~\cite{micuda2012noiseless}, and its counterintuitive ability to reduce the mean photon number of input states has now been demonstrated in several experiments~\cite{allevi2010subtraction,zhai2013pnr,bogdanov2017multiphoton,maganaloaiza2019multiphoton,katamadze2020multimode,nunn2022modifying}. As illustrated in the figure, an input state $\hat{\rho}_{in}$ with average photon number $\langle \hat{n} \rangle_{in}$ passes through a beamsplitter with variable reflectance $R$, and the output state is only accepted when no photons are reflected.  This is accomplished by actively ``heralding on zero'' photons (HoZ) using detector $D_{1}$~\cite{nunn2021heralding}, which occurs with a probability of success $P_{s}$ that decreases as $R$ is increased. Upon successful operation, the properties of the heralded output state $\hat{\rho}_{out}$ can then be measured with an auxiliary detection system represented by $D_{2}$.

It can be shown that this conditional measurement process implements the basic transformation $\ket{n}  \rightarrow t^{n} \ket{n}$ in the Fock state basis (here, transmittivity $|t|^{2} \equiv T = 1- R$), which results in an overall attenuation (i.e.,  $\langle \hat{n} \rangle_{out} < \langle \hat{n} \rangle_{in}$)  for all but Fock state inputs~\cite{gagatsos2014heralded}. Roughly speaking, this attenuation effect can be understood by considering that larger-$n$ terms in the Fock state expansion of a general input state are less likely to ``survive'' the HoZ process, resulting in a conditional re-weighting of the Fock state coefficients towards smaller-$n$ terms in the heralded output state.  Here we extend this idea to show how ZPS can also transform the statistical properties of input states, and then use these transformations to define a set of nonclassicality criteria based on ZPS measurements.

As a simple example of the ability of ZPS to perform statistical transformations, consider an input superposition state $\ket{\psi}_{in} = \frac{1}{\sqrt{2}}(\ket{1} + \ket{5})$, which has a mean photon number  $\langle \hat{n} \rangle_{in} = 3$ and a {\em positive} Mandel $Q$ parameter $Q_{in} \approx 0.33$, denoting super-Poissonian photon statistics. After ZPS with a standard 50:50 beamsplitter $(R =  \frac{1}{2})$, the output state becomes  $\ket{\psi}_{out} = \frac{4}{\sqrt{17}}(\ket{1} + \frac{1}{4}\ket{5})$, which has an attenuated mean photon number  $\langle \hat{n} \rangle_{out} \approx 1.2$ and a {\em negative} Mandel $Q$ parameter $Q_{out} \approx -0.28$. This negative value denotes sub-Poissonian photon statistics and thus non-classical properties of the heralded transmitted light~\cite{agarwal2012quantumoptics}.

Combined with a well-known ``no-go'' theorem which states that conditional photon-counting measurements at a beamsplitter cannot produce nonclassical output states unless the input is also nonclassical~\cite{ban1996statistics,kim2002entanglement}, this simple example shows how even a super-Poissonian input state may possess other ``hidden'' non-classical properties that could be revealed by ZPS measurements.  In this paper, we formalize this argument by characterizing the ZPS measurement process through a relative attenuation parameter $K(R)$ and deriving nonclassicality criteria based on its behavior. While $K(R)$ typically changes monotonically as beamsplitter reflectance $R$ is increased, we find that non-monotonic behavior arises for ``transformable'' states that change their sub- or super-Poissonian character, and that this ``transformability'' is always nonclassical.

The remainder of the paper is structured as follows: In Section~\ref{sec:atten} we define the ZPS relative attenuation parameter $K(R)$ and explore its relationship with $Q$. We illustrate the key features of $K(R)$ by comparing its behavior for three basic (non-transformable) states with two simple examples of transformable states. In Section~\ref{sec:noncl}, we extend the concept of ``transformability'' to define ZPS-based nonclassicality criteria, and in Section~\ref{sec:predict} we describe a method to predict which input states will statistically transform under ZPS.  In Section~\ref{sec:exam} we consider two more rich examples of ZPS input states-- the displaced squeezed state \cite{dodonov1994oscillations} and the catalyzed coherent state \cite{lvovsky2002catalysis}-- which illustrate the concept of transformability over a limited range of input state parameter space. Finally, in Section~\ref{sec:det} we consider the effects of realistic detectors on experimentally observing these ZPS-based statistical transformations.

\section{Relative attenuation parameter}\label{sec:atten}
 
When exactly zero photons are reflected by the beamsplitter in the setup of Fig. \ref{fig:overview}, ZPS is successful and {\em noiselessly attenuates} the input state  $\hat{\rho}_{in}$~\cite{micuda2012noiseless}. This process is described by the nonunitary operator $t^{\hat{n}}$, producing a modified state  $\hat{\rho}_{out}$ with resulting mean photon number:
\begin{equation}\label{eqn:nout}
    \langle \hat{n} \rangle_{out} = \frac{\sum_n p_n n T^n}{\sum_n p_n T^n},
\end{equation}
where $\{ p_{n} \}$ are the diagonal elements of $\hat{\rho}_{in}$ in the Fock basis. With the exception of pure Fock state inputs, this always results in $\langle \hat{n} \rangle_{out} < \langle \hat{n} \rangle_{in}$~\cite{gagatsos2014heralded}.  This can be seen by differentiating Eq.~\ref{eqn:nout} with respect to $T$:
\begin{equation}\label{eqn:dndt}
	\frac{d\langle \hat{n} \rangle_{out}}{dT} = \frac{1}{T}\left[\frac{\sum_n (n - \langle \hat{n} \rangle_{out})^2 p_nT^n}{\sum_n p_nT^n}\right] = \frac{\langle (\Delta n)^2\rangle_{out}}{T} \geq 0,
\end{equation}
 and noting that as $T$ decreases ($R$ increases), $ \langle \hat{n} \rangle_{out}$ is nonincreasing and only remains constant in the case of zero photon number variance $ \langle ( \Delta n)^2 \rangle_{out} = 0$.
 
We quantify this  photon number reduction through the relative attenuation parameter $K(R)$, which is defined as the ratio of the mean output photon number {\em with} HoZ (ZPS attenuation) to the mean output photon number {\em without} HoZ (``ordinary'' beamsplitter attenuation)~\cite{nunn2022modifying}:
 \begin{equation}
K(R) \equiv \frac{\langle \hat{n} \rangle_{out} }{(1-R) \langle \hat{n} \rangle_{in}}
 \label{eqn:kdef}
 \end{equation}

This particular definition facilitates the use of coherent states (Poissonian statistics) as a benchmark in experimental measurements. Because a coherent state impinging on the beamsplitter will produce two completely uncorrelated coherent states in the output ports~\cite{kim2002entanglement}, HoZ in the reflected port makes no difference on $\langle \hat{n} \rangle_{out}$ in the transmitted port. Consequently, $K(R) = 1$ for all $R$ for any coherent state input $\ket{\alpha}$.

In contrast, for any pure Fock state input $\ket{n}$ (sub-Poissonian statistics), ZPS will not reduce $\langle \hat{n} \rangle_{out}$ while ``ordinary'' beamsplitter attenuation grows with $R$, causing $K(R)$ to increase monotonically. Conversely, for any thermal state input  $\hat{\rho}_{th}$ (super-Poissonian statistics), ZPS will reduce $\langle \hat{n} \rangle_{out}$ {\em more} than ``ordinary'' beamsplitter attenuation~\cite{zhai2013pnr,nunn2022modifying}, causing $K(R)$ to decrease monotonically with $R$.

These trends in the behavior of $K(R)$ for different types of input state statistics can be formally seen by noting from Eq.~\ref{eqn:dndt} that the effects of ZPS are highly dependent on photon number variance, and by extension Mandel's $Q$ parameter, defined $Q = \frac{\langle ( \Delta n)^2 \rangle}{\langle \hat{n} \rangle} - 1$~\cite{mandel1979subpoissonian}. Combining the results of Eqs.~\ref{eqn:nout}-\ref{eqn:kdef}, one can obtain expressions for $Q$ in terms of the measurable quantity $K$:
\begin{equation}\label{eqn:qout}
    Q_{out}(R')=-\frac{(1-R')}{K(R')}\frac{dK}{dR}\Big|_{R=R'}
\end{equation}
for the output state, and for the input state $(R=0)$:
\begin{equation}\label{eqn:qin}
    Q_{in}=-\frac{dK}{dR}\Big|_{R=0}
\end{equation}
In both cases, the slope of $K(R)$ is determined by the sign of $Q$, with negative slopes denoting super-Poissonian statistics ($Q > 0$) and positive slopes denoting sub-Poissonian statistics ($Q < 0$). Consequently, the initial slope of $K(R)$  near $R = 0$ determines whether the input state was super- or sub-Poissonian~\cite{nunn2022modifying}, while any {\em changes} in the sign of the slope as $R$ is increased denote the statistical transformations of interest here.

\begin{figure}[t]
\includegraphics[scale=0.50,trim={210 60 220 40},clip]{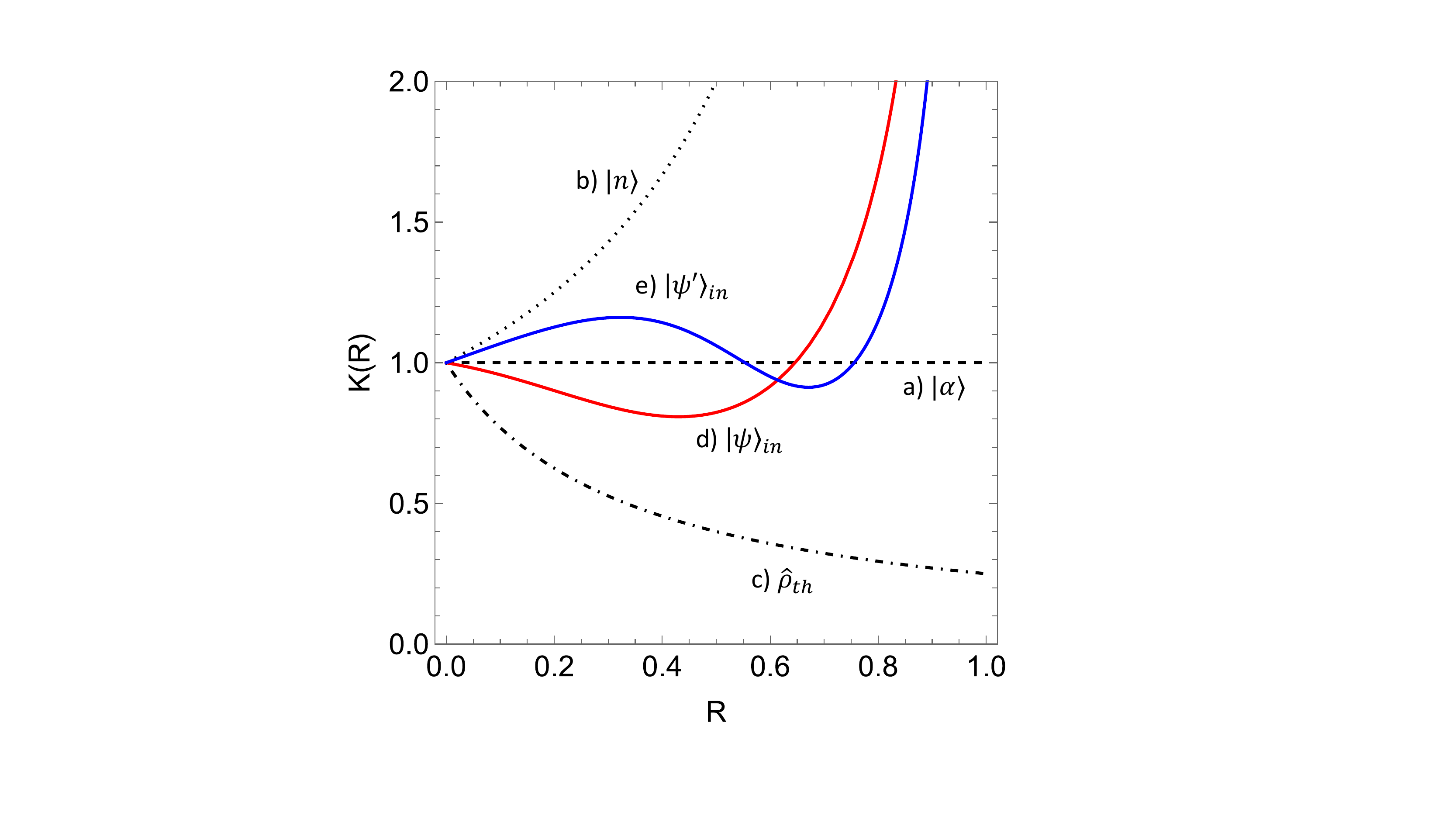}
\caption{Plots of the ZPS relative attenuation parameter $K(R)$ for five different input states. Non-monotonic behavior is a signature of a nonclassical input state, with local extrema denoting transformations between super- and sub-Poissonian photon statistics in the output state.  Plots (a) - (c) represent standard baseline examples of ``non-transformable'' input states, while plots (d) and (e) represent simple examples of states that are ``transformable'' by the ZPS process. For comparative purposes, $\langle \hat{n} \rangle_{in} = 3 $ in examples (a) - (c).}
\label{fig:fig2}
\end{figure} 

Figure~\ref{fig:fig2} illustrates these ideas by showing plots of relative attenuation $K(R)$ for five different input states. By definition, each state begins with $K=1$ at $R=0$ ($\langle \hat{n} \rangle_{out}=\langle \hat{n} \rangle_{in}$), and the plots diverge from this common point as reflectance increases. Plots (a) - (c) show the baseline examples of the three non-transformable states mentioned above ($|\alpha\rangle$,  $|n\rangle$, and  $\hat{\rho}_{th}$) which show monotonic behavior with initial slopes determined by Eq.~\ref{eqn:qin}. Plot (d) corresponds to  the transformable state described in Section~\ref{sec:intro}, $|\psi\rangle_{in} = \frac{1}{\sqrt{2}}(|1\rangle + |5\rangle)$, which shows non-monotonic behavior with a local minimum near $R \sim 0.4$. Past this point, we see the state becomes sub-Poissonian according to Eq.~\ref{eqn:qout}.

The super- to sub-Poissonian statistical transformation revealed by this non-monotonic behavior in $K(R)$ can be loosely understood in the following way:  when $R = 0$, the equal weighting of $|1\rangle$ and $|5\rangle$ in $|\psi\rangle_{in}$ yields a photon number variance larger than that of the coherent state benchmark (i.e., super-Poissonian), resulting in a negative initial slope for $K(R)$. As $R$ is increased, ZPS increasingly drives the equally weighted superposition towards the Fock state $|1\rangle$, causing a reduction in the photon number variance below that of the coherent state benchmark (i.e., sub-Poissonian) and thus a transition to a positive slope for $K(R)$.

Finally, plot (e) in Figure~\ref{fig:fig2} displays a reverse transformation from sub- to super-Poissonian statistics.  Here we use the input state $|\psi ' \rangle_{in} = \frac{1}{\sqrt{10}} |1\rangle +  \frac{3}{\sqrt{10}} |5\rangle$, which is heavily weighted towards $|5\rangle$ and thus starts with a small (sub-Poissonian) photon-number variance and  a positive initial slope for $K(R)$.  As $R$ is increased,  ZPS drives the state towards a more equally weighted superposition of $|1\rangle$ and $|5\rangle$, which has a larger (super-Poissonian) photon-number variance, and causes the transition to a negative slope for $K(R)$. As $R$ is further increased (past $R \sim 0.7$), we replicate the behavior seen in plot (d): ZPS drives the superposition closer to the Fock state $|1\rangle$ and $K(R)$ once again has a positive slope. Overall, this state undergoes two statistical transformations-- from sub- to super-Poissonian statistics and back again-- over the full range of $R$.

 \section{Nonclassicality criteria}\label{sec:noncl}
 
It is well known that conditional measurements at a beamsplitter (such as ZPS) cannot produce a nonclassical output unless the input is also nonclassical~\cite{ban1996statistics,kim2002entanglement}. Here, {\em nonclassical} states are defined in the usual way as those which cannot be expressed as a mixture of coherent states~\cite{dodonov2002nonclassical}. The Glauber-Sudarshan $P$-representation for such states cannot be interpreted as a valid probability distribution, becoming highly singular or taking on negative values \cite{glauber1963coherence,sudarshan1963equivalence}. It is generally accepted that sub-Poissonian statistics are nonclassical by this definition \cite{agarwal2012quantumoptics,dodonov2002nonclassical}. It follows, therefore, that any input state which {\em becomes} sub-Poissonian after ZPS must have been nonclassical to begin with.

This useful restriction can be formulated in terms of the relative attenuation parameter $K(R)$. Any classical (i.e., not nonclassical) input state $\hat{\rho}_{in}$ is bound by the following inequality:
 \begin{equation}
\frac{dK}{dR} \leq 0  \quad\text{for all $R$} \label{eqn:dkdrlim}
 \end{equation}

 Any state which is sub-Poissonian before or after ZPS will violate the above inequality and certify $\hat{\rho}_{in}$ as nonclassical. Given that $K = 1$ at $R = 0$ by definition, it immediately follows that for classical states:
 \begin{equation}
K(R) \leq 1  \quad\text{for all $R$} \label{eqn:klim}
 \end{equation}
As seen in Figure~\ref{fig:fig2}(a), the coherent state input $|\alpha\rangle$ saturates these bounds with $K(R) = 1$ and $dK/dR = 0$ for all $R$.

To summarize, violations of Eqs.~\ref{eqn:dkdrlim} and/or \ref{eqn:klim} represent nonclassicality criteria that can be easily measured with the ZPS setup of Fig.~\ref{fig:overview}, and are based on the ability of ZPS to generate sub-Poissonian statistics in the output. This transformation occurs \textit{if and only if} the slope of $K(R)$ is positive for some value of $R$. Alternatively, a single measurement of $K(R)>1$ for some $R$ is sufficient (though not necessary) to identify such a nonclassical state. Of course, sub-Poissonian input states are guaranteed to trigger both of these nonclassicality criteria.

 \section{Predicting transformability}\label{sec:predict}
 
From the simple examples in Section~\ref{sec:atten}, it is clear that not all super-Poissonian states can transform into sub-Poissonian states under ZPS, or vice versa. We have shown that ``transformable'' states must be nonclassical in Sec.~\ref{sec:noncl}, but the nonclassicality criteria in Eqs.~\ref{eqn:dkdrlim} and \ref{eqn:klim} provide little {\em a priori} insight into which states will actually transform. In this section, we derive sufficient conditions for predicting transformability based on the first few terms of the photon number distribution.

The key insight is that if a state only transforms \textit{once} over the full range of $R$, we need only to consider the behavior of $K(R)$ at maximum attenuation to verify if the statistics have changed. In the limit $R\rightarrow 1$, $K(R)$ and its derivative are given by:
\begin{align}
    &\lim_{R\to 1}K(R) = \frac{1}{\langle \hat{n}\rangle_{in}}\left(\frac{p_1}{ p_0}\right) \label{eqn:kend}\\
    &\lim_{R\to 1}\frac{dK}{dR} = \frac{1}{\langle \hat{n}\rangle_{in}}\left(\frac{p_1^2 - 2p_0p_2}{p_0^2}\right) \label{eqn:dkdrend}
\end{align}
Remarkably, these values are entirely determined by the mean photon number $\langle \hat{n}\rangle_{in}$ and first three photon number probabilities $(p_0, p_1, p_2)$ of the input state.

For a super-Poissonian input state, transformability amounts to violating the classical bounds in Eqs.~\ref{eqn:dkdrlim} and \ref{eqn:klim}, meaning $dK/dR>0$ or $K>1$ for some value of $R$. Combining this with Eqs.~\ref{eqn:kend} and \ref{eqn:dkdrend}, we obtain the following transformability criteria:
 \begin{align}
 p_0 &= 0 \label{eqn:p0lee}\\
p_0\langle\hat{n}\rangle_{in} &< p_1 \label{eqn:p0rule}\\
2p_0p_2 &< p_1^2\label{eqn:klyshko}
 \end{align}
Satisfying any of the above is sufficient to show that a super-Poissonian input state will transform for sufficiently large $R$.

We note that Eq.~\ref{eqn:p0lee} is equivalent to Lee's theorem, which states that any input state with zero vacuum probability $p_{0} = 0$ is nonclassical~\cite{lee1995theorem}, and is often invoked to understand the nonclassical nature of photon-added states \cite{agarwal1991nonclassical,barnett2018statistics}. In the case of ZPS, it is clear that if $p_{0}= 0$, $K(R)$ and its derivative will diverge to infinity as $R \rightarrow 1$, triggering both nonclassicality criteria in Eqs.~\ref{eqn:dkdrlim} and \ref{eqn:klim}. In addition, Eq.~\ref{eqn:klyshko} is equivalent to Klyshko's well-known nonclassicality criterion~\cite{klyshko1996observable}. Thus, all super-Poissonian states which trigger Lee's or Klyshko's nonclassicality criteria are transformable.

Conversely, a sub-Poissonian input state will transform if $dK/dR<0$ or $K<1$ for some value of $R$. It follows that satisfying either of Eqs.~\ref{eqn:p0rule} or \ref{eqn:klyshko}, with the inequalities reversed, is sufficient to show that a sub-Poissonian state is transformable. Lee's criteria (Eq.~\ref{eqn:p0lee}) cannot be used to predict transformability for sub-Poissonian states.

The above methods for predicting transformability rely only on comparing statistics at $R=0$ and $R\rightarrow1$. Thus, states which transform an \textit{even} number of times over the full range of $R$ may not trigger these criteria, as in the example of Figure~\ref{fig:fig2}(e) which is sub-Poissonian at both extremes. This also holds for super-Poissonian input states, so satisfying Klyshko's inequality (Eq.~\ref{eqn:klyshko}) is sufficient but not necessary to be transformable. For example, any mixture/superposition of 0, 2 and 6 photons does not satisfy Klyshko's inequality, yet it can be shown that the specific case of $p_0=0.04$, $p_2=0.48$, $p_6=0.48$ corresponds to a transformable super-Poissonian state.

 \section{Input state parameters}\label{sec:exam}
 
For many types of quantum optical states, the initial statistical character is determined by a set of input parameters. A trivial example is the two-term Fock state superposition considered in Figs. 2(d) and (e), $|\psi \rangle_{in} = \gamma |1\rangle + \sqrt{(1 - \gamma^{2})} |5 \rangle$, which was initially super- or sub-Poissonian based on the value of $\gamma$.  More complex examples include squeezed and/or displaced states, in which the magnitude and angle of squeezing and displacement can result in dramatically different photon statistics~\cite{deoliveira1990properties,grosse2007antibunching}, and conditional state preparation techniques like photon catalysis~\cite{bartley2012multiphoton}.
 
It is interesting to extend this idea to address the following question: for a given type of input state, what regions of its input parameter space will enable it to transform under ZPS?  Here we consider two rich examples within this context-- displaced squeezed states, as considered by Dodonov {\em et al.}~\cite{dodonov1994oscillations},  and catalyzed coherent states~\cite{lvovsky2002catalysis}.

\subsection{Displaced squeezed state}
Noiseless attenuation of general gaussian states was previously considered in Ref.~\cite{gagatsos2014heralded}, and it was shown that ZPS preserves gaussianity. However, we show here that ZPS may not preserve the sub- or super-Poissonian character of these states for a significant portion of their parameter space. We consider a displaced squeezed state with real squeezing parameter $r$ and displacement magnitude $z$. To restrict our space to two variables, the angles of squeezing and displacement are chosen so the two quadrature variances are equal $\sigma_x=\sigma_p$, and their means opposite $\langle \hat{x} \rangle=-\langle \hat{p} \rangle$, following Dodonov {\em et al.}~\cite{dodonov1994oscillations}.

\begin{figure}[b]
    \centering
    \includegraphics[scale=0.36,trim={130 30 140 50},clip]{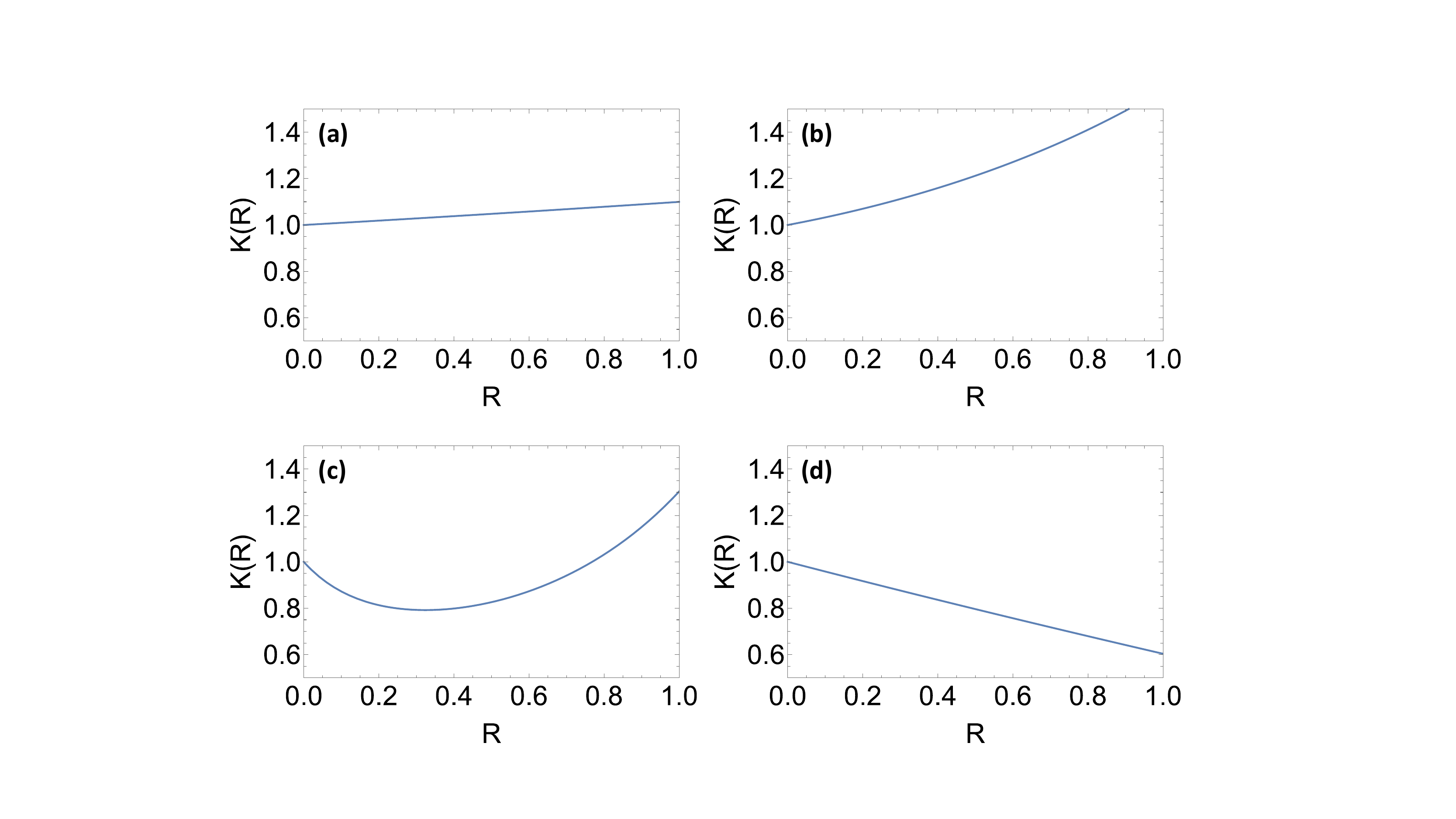}
    \caption{Relative attenuation $K(R)$ for a displaced squeezed state with parameters (a) $z=1$, $r=0.05$, (b) $z=1$, $r=0.35$, (c) $z=1$, $r=1$, and (d) $z=0.1$, $r=0.1$. Cases (a), (b) and (d) do not exhibit statistical transformations, while case (c) illustrates a transformation from super- to sub-Poissonian statistics under ZPS.}
    \label{fig:SPEC}
\end{figure}

Starting with the photon statistics derived in~\cite{dodonov1994oscillations}, the relative attenuation $K(R)$ can be shown to be:
    \begin{equation}
    \begin{split}
       K(R)= \frac{1}{\sinh^2{r} + 
  z^2} \left[\frac{1}{2\left(\coth{r} - (1 - R)\right)}\right. 
  \\- 
   \left.\frac{1}{\coth{r} +\left( 1 - 
     R\right)}\left( \frac{1}{2} - 2 z^2 \frac{\coth{2r} + 1}{1 + (1 - R)\tanh{r}} \right)\right]
     \end{split}
    \end{equation}
Figure~\ref{fig:SPEC} shows plots of $K(R)$ for multiple combinations of displacement and squeezing parameters $z$ and $r$. For $z=1$ and $r=1$ in Fig.~\ref{fig:SPEC}(c), a local minimum appears. This indicates that after sufficient noiseless attenuation, the super-Poissonian input state undergoes a transition to sub-Poissonian statistics in the output.

\begin{figure}
    \centering
    \includegraphics[scale=0.41,trim={165 120 150 120},clip]{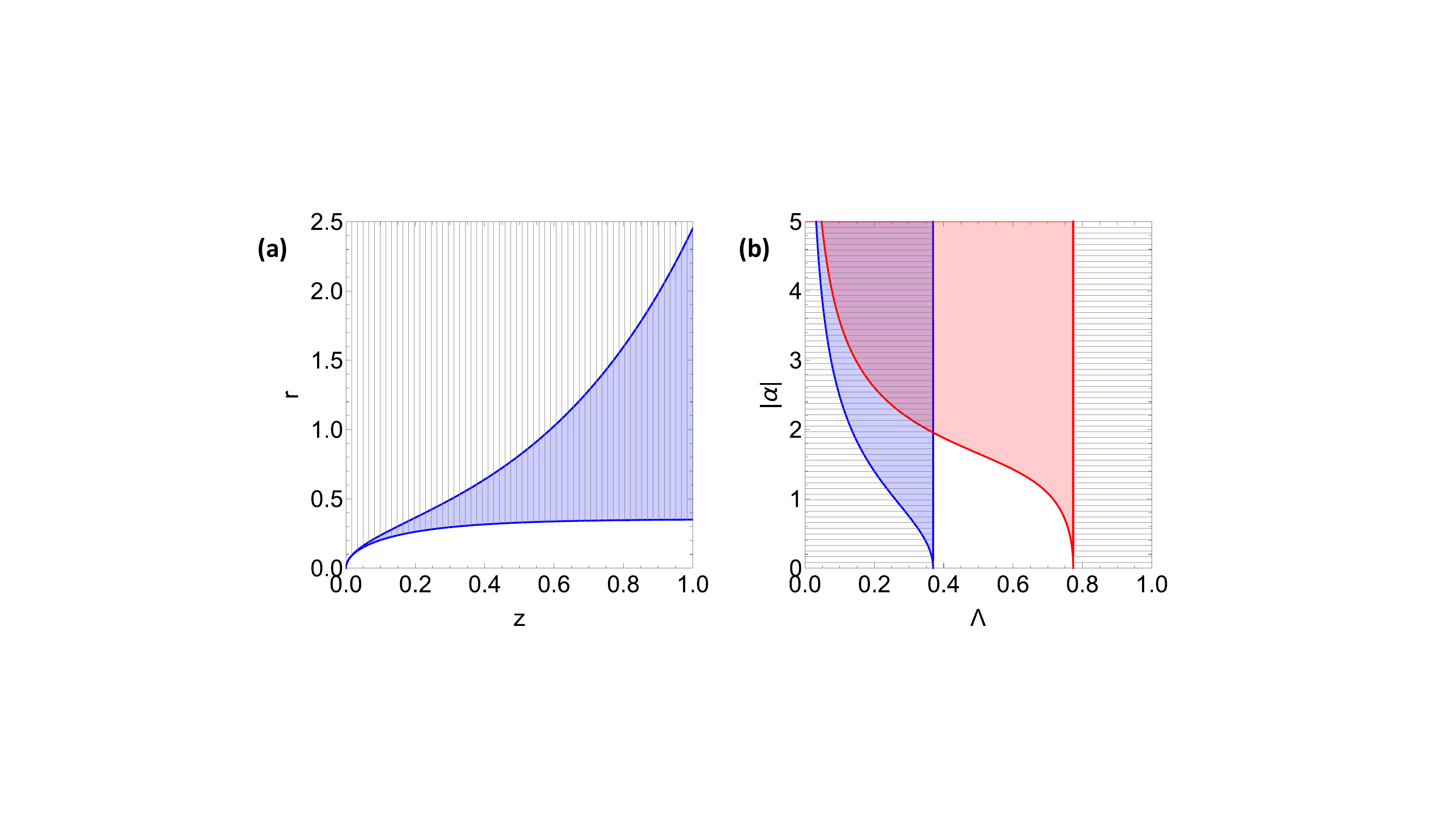}
    \caption{Regions shaded blue and red correspond to parameter spaces for which the input of (a) a displaced squeezed state and (b) a catalyzed coherent state, has a local minimum and maximum respectively in $K(R)$. For (b), these regions overlap. An example from within the overlapping region is shown in Fig. ~\ref{fig:CCS}(c). }
    \label{fig:regions}
\end{figure}

There is a wide portion of parameter space for which this transition is possible, as shown by the blue shaded region in Fig.~\ref{fig:regions}(a). The bounding curves are calculated analytically in Mathematica. The bottom curve corresponds to the bound of Klyshko's inequality (Eq.~\ref{eqn:klyshko}), while the top curve divides sub- from super-Poissonian input states. This leaves three total regions: (i) the blank bottom-most region which contains super-Poissonian states with low squeezing, including squeezed vacuum states; (ii) the transformable super-Poissonian states, with moderate displacement and squeezing; and (iii) the upper-most region of sub-Poissonian states, including those with the highest squeezing. There are no cases in which ZPS transforms a sub-Poissonian state of this kind into a super-Poissonian one.

As discussed in Sec.~\ref{sec:noncl}, the transition in Fig.~\ref{fig:SPEC}(c) is clearly nonclassical according to Eq.~\ref{eqn:dkdrlim}, and likewise satisfies Klyshko's inequality (Eq.~\ref{eqn:klyshko}). However, it is interesting to note that the input has no ``hidden'' higher-order sub-Poissonian statistics as defined in Ref.~\cite{erenso2002higherorder}. In other words, it can be shown that the normalized correlations functions for this state satisfy $g^{(n)}(0)>1$ for all $n$. In this sense, ZPS does not ``reveal'' sub-Poissonian statistics, but rather generates them from other nonclassical aspects of the initial number distribution.

\subsection{Catalyzed coherent state}
Catalyzed coherent states (CCS) can be generated by mixing a coherent state $\ket{\alpha}$ with a single-photon Fock state at a beamsplitter with reflectance $\Lambda$, and conditioning the output on the detection of exactly one photon in the auxiliary output mode~\cite{lvovsky2002catalysis}. As first proposed by Xu and Yuan, a similar catalysis procedure can be implemented with the signal and idler modes of an optical parametric amplifier (OPA) with an equivalent catalysis parameter $\Lambda=1-1/g^2$, where $g$ is the gain of the OPA~\cite{xu2016catalysis,shringarpure2019generating,barnett1999equivalence}.

Here, we consider an input CCS with coherent state amplitude $\alpha$ and catalysis parameter $\Lambda$. It can be shown that the photon number distribution for this state is:
\begin{figure}%
    \centering  %
    \includegraphics[scale=0.36,trim={130 30 140 50},clip]{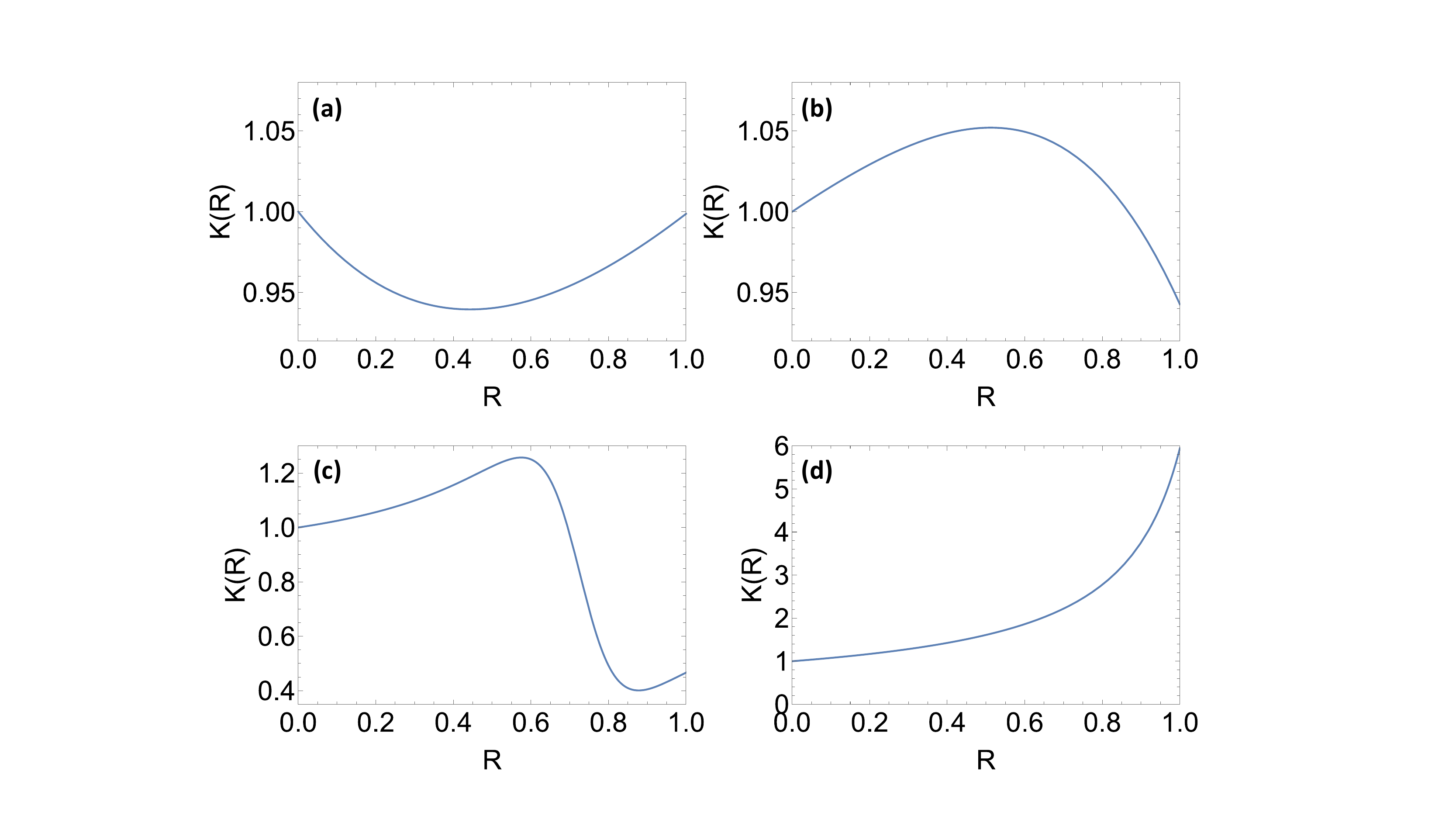}
    \caption{Relative attenuation $K(R)$ for a catalyzed coherent state with parameters (a) $\Lambda=0.3$, $\alpha=1$, (b) $\Lambda=0.75$, $\alpha=1$, (c) $\Lambda=0.2$, $\alpha=4$, and (d) $\Lambda=0.9$, $\alpha=1$. Cases (a) and (b) exhibit a super-to-sub- and sub-to-super-Poissonian transformation, respectively, while case (c) shows both and (d) shows neither.\label{fig:CCS}}
    \end{figure} 
    
    \begin{equation}
        p_n=\frac{e^{-|\alpha| ^2 (1-\Lambda)} |\alpha| ^{2 n} (1-\Lambda )^{n-1} (\Lambda +\Lambda  n-1)^2}{n! \left[1-\Lambda-|\alpha |^2 \Lambda  (\Lambda^2 -4 \Lambda +2) \right]}
    \end{equation}
Using Eqs.~\ref{eqn:nout} and \ref{eqn:kdef}, $K(R)$ is then given by:
\begin{equation}
	K(R)\propto\frac{\splitdfrac{1-4\Lambda(1-\Lambda)-\alpha^2\Lambda(2-5\Lambda)(1-R)}{+\alpha^4\Lambda^2(1-\Lambda)^2(1-R)^2}}{\splitdfrac{1-\Lambda-\alpha^2\Lambda(2-3\Lambda)(1-R)}{+\alpha^4\Lambda^2(1-\Lambda)(1-R)^2}}
\end{equation}
with the necessary normalization such that $K=1$ at $R=0$.

Like the displaced squeezed state, relative attenuation of the CCS can exhibit a local minimum for particular values of the input parameters. In Fig.~\ref{fig:CCS}(a), a minimum is found for $\Lambda=0.3$ and $\alpha=1$. Just like the example in Fig.~\ref{fig:SPEC}(c), this state is super-Poissonian to arbitrarily high order ($g^{(n)}(0)>1$) and satisfies Klyshko's nonclassicality criteria. Unlike the previous example, however, $K(R)$ can also have a local \textit{maximum} for different parameters, as in Fig.~\ref{fig:CCS}(b). In this example, all higher-order statistics of the input state are sub-Poissonian ($g^{(n)}(0)<1$), and yet attenuation by ZPS produces super-Poissonian output statistics for $R\gtrsim 0.52$.

The parameter space for local minima and maxima is shown in Fig.~\ref{fig:regions}(b), again calculated analytically with Mathematica. Interestingly, the two regions overlap, allowing for input cases with both a minimum and maximum, as in Fig.~\ref{fig:CCS}(c). Additionally, Klyshko's inequality is satisfied when $\Lambda\lesssim0.365$ and when $\Lambda\gtrsim0.775$. The former region contains all super-to-sub transformations, while the latter contains no transforming states.

 \section{Realistic detectors}\label{sec:det}
 
In a real experiment, the ability to observe the statistical transformations of interest will critically depend on the performance of detectors $D_{1}$ and $D_{2}$ in Fig.~\ref{fig:overview}. Here we consider the effects of dark counts and inefficiency in both detectors, as well as the lack of photon number resolution (PNR) for $D_2$ in the transmitted output port of the beamsplitter.

If heralding detector $D_1$ has a reduced effective efficiency $\eta_1$ (including losses in the reflected mode or before the beamsplitter) then it can be shown that~\cite{nunn2022modifying}:
\begin{align}
	K_{\text{exp}}(R)= K_{\text{ideal}}(R \eta_1)\label{eqn:keff}
\end{align}
If $\eta_1$ is known, then $K_{\text{ideal}}(R)$ can be recovered over the smaller domain $R\in[0,\eta_1)$. However, if any statistical transformations or other interesting behavior only exist beyond $R=\eta_1$, they can no longer be generated in the transmitted output port. In the limit of $\eta_1\rightarrow0$, ZPS is equivalent to ordinary attenuation, which can never change the sign of Mandel's $Q$ parameter~\cite{alleaume2004singlephoton}.

\begin{figure}
    \centering
    \includegraphics[scale=0.46,trim={310 150 310 150},clip]{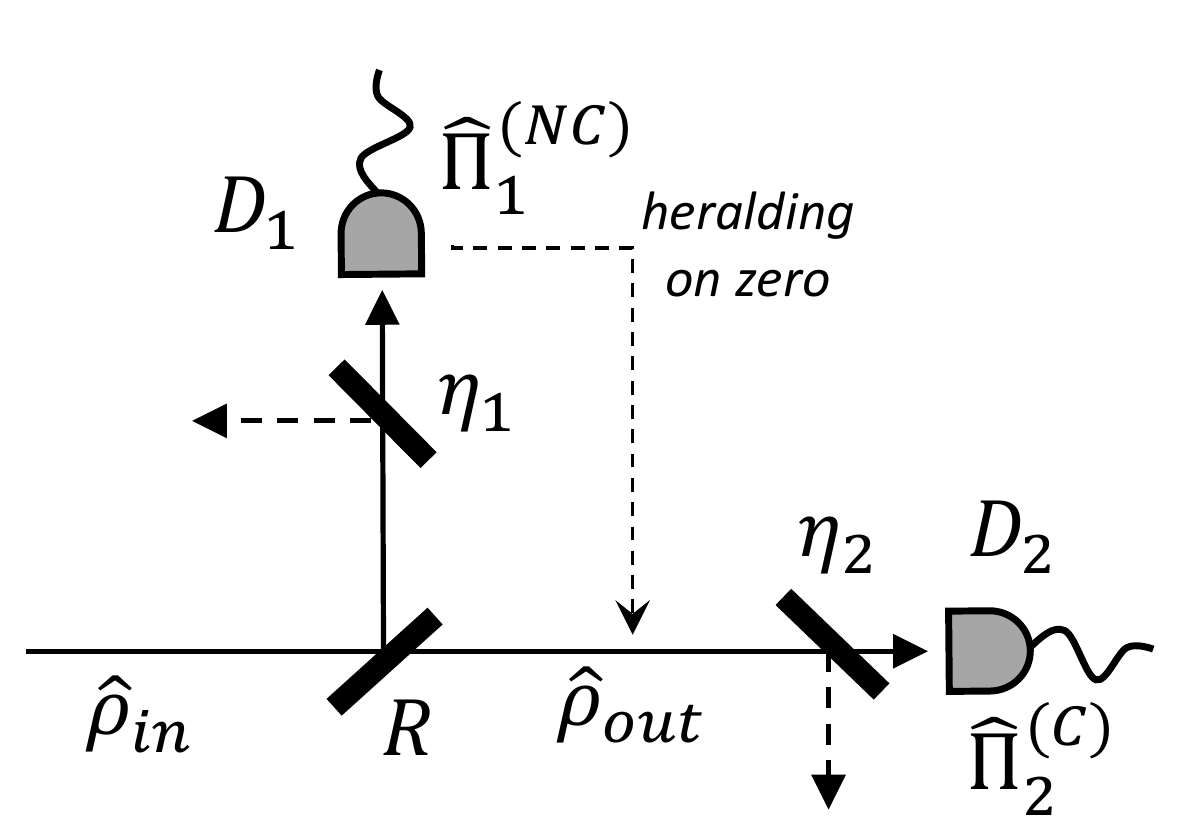}
    \caption{ZPS measurement with two non-PNR detectors. Detector $D_2$ measures the rate of clicks with and without conditioning on a no-click at $D_1$. Detector inefficiencies $\eta_1$ and $\eta_2$ are modeled with beamsplitters that reflect photons into the environment. To approximate $K(R)$ without PNR, the effective efficiency of $D_2$ must satisfy $|\eta_2 n|\ll1$.}
    \label{fig:detectors}
\end{figure}

Meanwhile, $D_2$ efficiency $\eta_2$ or loss in the output mode has no effect on $K(R)$ as defined in Eq.~\ref{eqn:kdef}. The numerator and denominator of this ratio would be reduced by the same efficiency factor $\eta_2$, which then cancels. However, the efficiency $\eta_2$ must be carefully considered if using a non-PNR detector $D_2$ (see Fig.~\ref{fig:detectors}). In this case, mean intensity or expected photon number at $D_2$ is replaced with the probability of a click event:
\begin{align}\label{eqn:kclick}
    K_{\text{click}}(R)&\equiv
    \frac{P(C_2|NC_1)}{P(C_2)} \nonumber \\
    &= \frac{
     \text{tr}\left\{\hat{B}\hat\rho_{in}\hat{B}^\dagger\hat\Pi^{(NC)}_1\hat\Pi^{(C)}_2\right\}}
     {\text{tr}\left\{\hat{B}\hat\rho_{in}\hat{B}^\dagger\hat\Pi^{(C)}_2\right\}\text{tr}\left\{\hat{B}\hat\rho_{in}\hat{B}^\dagger\hat\Pi^{(NC)}_1\right\}} \nonumber \\
     &=	\frac{\sum_np_n\left[(1-R\eta_1)^n-(1-R\eta_1-T\eta_2)^n\right]}{\left[\sum_np_n(1-R\eta_1)^n\right]\left[1-\sum_np_n(1-T\eta_2)^n\right]}
\end{align}
where the subscripts $1$ and $2$ indicate detection channels $D_1$ and $D_2$, $\hat{B}$ is the unitary beam-splitter operator~\cite{agarwal2012quantumoptics}, and we have used the standard POVMs for non-PNR detectors for ``click'' $(C)$ and ``no-click'' $(NC)$ events~\cite{stevens2013photon}:
\begin{align}
	\hat\Pi^{(C)}_i &= \mathbbm{1} - \Pi^{(NC)}_i \nonumber \\
	\hat\Pi^{(NC)}_i &= \sum_n (1-\eta_i)^n\ket{n}\bra{n}
\end{align}
for $i=1,2$. We find that $\lim_{\eta_2 \to 0}K_{\text{click}}(\eta_1,\eta_2,R)=K(\eta_1R)$ after applying L'H\^{o}pital's rule. Alternatively, the binomial approximation can be applied so long as $|\eta_2 n|\ll1$ and higher-number terms are insignificant. This is similar to the requirements for measuring normalized correlation functions $g^{(n)}$. In these multiphoton coincidence measurements, non-PNR detectors are effective for low efficiency and low counting rates, when the detector response to photon number is approximately linear~\cite{avenhaus2010higherorder}. These measurements are considered loss-tolerant exactly like the transmitted mode in ZPS.

Dark counts at the heralding detector $D_1$ have no effect on $K(R)$, only reducing the probability of success for HoZ~\cite{nunn2021heralding}. These false click events can be safely ignored so long as $D_1$ dark counts are completely uncorrelated with the counts at the other detector $D_2$.

In contrast, dark counts at $D_2$ do have an effect on the measured relative attenuation. For some dark count probability $d$ at $D_2$, Eq.~\ref{eqn:kclick} is modified as follows:
\begin{equation}
    K_{\text{dark}}(R)=\frac{P(C_2|NC_1)+d\cdot P(NC_2|NC_1)}{P(C_2)+d\cdot P(NC_2)}
\end{equation}
If we assume the probability of a simultaneous dark and ``true'' count is negligible-- i.e., counting rates are sufficiently low with and without HoZ, then this becomes:
\begin{equation}\label{eqn:kdark}
     K_{\text{dark}}(R)\approx\frac{P(C_2|NC_1)+d}{P(C_2)+d}
\end{equation}
If the dark count rate $d$ is simply subtracted from both the numerator and denominator, we regain the original ratio $K_{\text{click}}$. In any case, lower $D_2$ dark counts are desirable to improve the signal-to-noise ratio.

In summary, the function $K(R)$ can be reliably measured with currently available single-photon detectors, even without PNR capability. The heralding detector $D_1$ requires high efficiency to generate the full range of possible output statistics, but is robust to dark counts and other uncorrelated background noise. The output detector $D_2$ has essentially opposite requirements, requiring low effective efficiency (if non-PNR) and relatively low dark counts.

\section{Summary \& Conclusions}\label{sec:con}

We have shown how the conditional measurement process of zero-photon subtraction (ZPS) can transform certain super-Poissonian states into sub-Poissonian states, and vice versa. These effects can be experimentally observed by measuring the relative attenuation parameter $K(R)$, which is simply defined as the ratio of ZPS-based attenuation to ordinary beamsplitter attenuation. Because the slope of $K(R)$ is proportional to Mandel’s $Q$-parameter, an observed local minimum or maximum as the beamsplitter  reflectivity is tuned from $R=0\rightarrow1$ is the signature of a statistical transformation of interest.

We described how input states which transform in this way are necessarily nonclassical, which allowed us to establish nonclassicality criteria based on ZPS measurements. The connection between these criteria and Klyshko and Lee’s theorems~\cite{klyshko1996observable,lee1995theorem} showed that certain restrictions on photon number probabilities provide sufficient, but not necessary, conditions to predict \textit{a priori} which input states will transform through the ZPS process.

We considered several simple examples of input states that illustrated the basic physics of these ZPS-based statistical transformations, as well as two more complex example states that showed interesting parameter-dependent behavior. In all cases, the effects of realistic detector parameters (low efficiency and dark counts) were found to only moderately reduce the ability to observe the statistical transformations of interest. Consequently, an experimental demonstration could be feasible with currently available detector technology~\cite{eisaman2011detectors}, although high-fidelity preparation of the various quantum optical input states considered here would remain a challenge.

\section*{Acknowledgements}
We would like to thank J. D. Franson for many valuable discussions pertaining to this research. This work was supported by the National Science Foundation under Grant No. 2013464.

%

\end{document}